\documentstyle [12pt] {article}
\newcommand{\cs}[3]{{{#3} \brace {#1 #2}}}

\begin{document}

\title{SN1987A:  Temporal Models}
\author{M.I.Wanas, M.Melek and M.E.Kahil \\ 
 Astronomy Department, Faculty of Science,
Cairo University, Giza, Egypt\\
  E-mail: wanas@mailer.scu.eun.eg}
\maketitle
 \begin{abstract}{ It is well known that carriers of
astrophysical information
 are massless spinning particles. These carriers are photons,
 neutrinos and, expectedly, gravitons. All these particles are
 emitted during supernova events. Information carried by these
 particles characterize their sources, but such information are affected by
 the trajectories of the carriers. Recently, it is shown that
 these trajectories are spin dependent. Knowing these trajectories
 and the arrival times of such particles to the detectors, a
 spin dependent model is constructed and compared with the
 conventional spin independent model.}
\end{abstract}

\section{Introduction}
   To construct a chronological model for a supernova (SN), i.e. a model arranging
main events on a certain time scale, there are at least two
different methods:\\ 1- In the first method one needs theories
investigating supernovae mechanism . Such theories depend on some
suggested theoretical assumptions and certain parameters. Once we
insert the parameters (characterizing a certain supernova) in the
theory then we get a temporal ( chronological) model for this
supernova.\\ 2- The input required for this method contains both
observational and theoretical information, these are :\\ (i) The
exact arrival times of the carriers of astrophysical information
coming from a certain SN to the Earth's detectors. \\
  (ii) The (suggested) trajectories followed by these carriers from this SN to
the Earth's detectors. \\ Having these information, one can
construct a temporal model for the SN.

Carriers of astrophysical information are mainly photons of
different frequencies. Recently, astronomers started to extract
astrophysical information from a second type carriers, the
neutrino. We expect, in the near future, to be able to extract
astrophysical information from a third type of carriers, the
graviton. Fortunately, the three types of carriers are assumed to
be emitted from SN events. Photons and neutrinos were detected
from SN1987A, and there is a claim that gravitons were also
detected.

   In  February $23^{rd}$ 1987 a supernova of type II in the Large Magallanic
Cloud (LMC) has been observed (cf. [1]). Observations have shown
that the arrival time of neutrinos at the Kamiokande detectors has
been in: Feb. $23^{rd}$, $7^h$ $35^m$ UT; while the arrival time
of photons has been in: Feb. $23^{rd}$, $10^h$ $40^m$ UT. Thus
neutrinos have arrived  $3^h$ $5^m$ earlier than photons. The bar
gravitational waves antennae in Rome and Maryland recorded
relatively large pulses which arrived about 1.2 sec. earlier than
neutrinos. The probability that the correlation, between
gravitational waves antennae and neutrino detectors, being
accidental is found to be about $10^{-6}$ (cf. [2] and [3]) .

   In the present work we will construct two temporal models for SN1987A using
the recorded times of arrival of the three carriers of
astrophysical information. The first model depends on the
assumption that the trajectory of massless particle (spinning or
spinless) in a background gravitational field, is a null-geodesic.
The second model depends on the assumption that the trajectory of
a massless particle, in a gravitational field, is spin dependent.
In section 2 we give a brief review of the null-geodesic equation
and the spin dependent equation. In section 3 we construct the two
temporal models for SN1987A. In section 4 we discuss and compare
the two models in view of theories investigating supernovae
mechanism. \\ \\ \\

\section{Trajectories of Massless Spinning Particles in Gravitational Fields}

 In the context of general relativity, the trajectory of massless particles
in a background gravitational field is given by the null-geodesic
equation,

\begin{equation}
{{\frac{dU^\mu}{d\lambda}} + \cs{\nu}{\sigma}{\mu}\ U^\nu U^\sigma
= 0},
\end{equation}
where $U^\mu$ is the null tangent to the trajectory and $\lambda$
is a parameter varying along the trajectory,$
\cs{\nu}{\sigma}{\mu}\ $ is the Christoffel symbol characterizing
the background gravitational field. The field is obtained as a
solution of the field equations of GR or any other field theory
written in Riemannian geometry.

   Recently, new path  equations, in  the absolute parallelism (AP) geometry, which can
be considered as  generalization of the paths of Riemannian
geometry (geodesic and null-geodesic) are obtained [4]. The
generalization of these equations can be written in the form [5],

\begin{equation}
{{\frac{dZ^\mu}{d\tau}} + \cs{\nu}{\sigma}{\mu}\ Z^\nu Z^\sigma =
- {\frac{n}{2}} \alpha \gamma \Lambda_{(\nu \sigma)}.^\mu~~  Z^\nu Z^\sigma}
\end{equation}
 where $Z^\mu$ is the tangent to the path, $\Lambda^\mu_{. \nu \sigma}$
is the torsion of space-time produced by the background
gravitational field, $(\alpha)$ is the fine structure constant,
$(\gamma)$ is a numerical parameter of order unity and $(n)$ is a
natural number taking the value 0, 1, 2,... for 0,
${\frac{1}{2}}$, 1,...spinning particles, respectively. The
brackets ( ) are used, in the equation, for symmetrization. The
term on the R. H. S. of (2) is suggested to represent a type of
interaction between the spin of the moving particle and the
torsion generated by the background gravitational field. Equation
(2) can be used to study the trajectory of a massless test
particle in a background gravitational field. The field is
obtained as a solution of the field equations any field theory
(including GR) written in AP-spaces.

Assuming that the background gravitational field is weak and
static, then linearizing (2) the following relation is obtained
[5],

\begin{equation}
{\Phi_S = (1 - {\frac{n}{2}} \alpha \gamma) \Phi_N}
\end{equation}
where $\Phi_s$ is the gravitational potential felt by a spinning
particle, and $\Phi_N$ is the Newtonian potential. If the particle
is a spinless one $(n = 0)$, then the two potentials will
coincide. It is to be noted that, in this case equation (2) will
be reduced either to geodesic equation, or to the null-geodesic
equation (1) upon reparameterization.

\section{Temporal Models}
\subsection { {\bf Spin Independent Model}}

It is well known that the time interval required for a photon  to
traverse a given distance is larger in the presence of a
gravitational field having a potential $\Phi(r)$. The time delay
is given by (cf. [6]),

\begin{equation}
{\Delta t_{GR} = const.~~ \int_e^a \Phi(r) dt}
\end{equation}
where (e) and (a) are the times of emission and absorption of the
photon, respectively. If the field is weak and static, $\Phi$ is
taken to denote the Newtonian gravitational potential $\Phi_N$.
Authors usually use (4), to get the time delay, for both photons
and neutrinos, and it could also be applied for gravitons. In the
SN1987A time delay calculations, $\Phi(r)$ is taken to be the
Newtonian gravitational potential of the Milky Way galaxy (cf.
[6],[7] and [8]).

 If the three carriers are moving along the same
trajectory (null-geodesic), then all will encounter the same
delay. Consequently, the difference in the arrival times will
coincide with the difference in the emission times at the source
of these carriers, SN1987A. The model is now clear. Neutrinos,
signaling the collapse of the core, are emitted at the same time
with gravitons indicating a sudden change of symmetries of
space-time. After $ 3^h~~5^m$ photons, indicating the maximum
brightness of the envelope, are emitted. This model is discussed
in the next section. \\
\subsection {{\bf Spin Dependent Model}}

It is obvious from (3) that $\Phi_S$ is different from $\Phi_N$,
i.e. the gravitational potential felt by a spinning particle
slightly differs from that felt by a spinless particle, due to the
new suggested interaction mentioned above. Thus we are going to
replace $\Phi(r)$ in (4) by $\Phi_S$ given by (3). This is done to
account for the suggested interaction between the spin of moving
particle and torsion of the background space-time. In this case we
get,

\begin{equation}
{\Delta t_s = (1 - {\frac{n}{2}} \alpha \gamma) \Delta t_{GR} ,}
\end{equation}
the index $s$ indicates that the time delay is calculated taking
into account the quantum spin of the moving particle, while GR
refers to general relativity. It is clear from (5) that $\Delta
t_s$ will not be the same for neutrinos $(n = 1)$, for photons $(n
= 2)$ and for gravitons $(n = 4)$. Thus, particles with different
spin will encounter different time delay, i.e.

\begin{equation}
{(\Delta t_s)_\gamma = (1 - \alpha \gamma) \Delta t_{GR}
~~~~for~~a~ photon,}
\end{equation}

\begin{equation}
{(\Delta t_s)_\nu = (1 - {\frac{1}{2}} \alpha \gamma) \Delta
t_{GR}~~~~for~a~ neutrino,}
\end{equation}

\begin{equation}
{(\Delta t_s)_g = (1 - 2~ \alpha \gamma) \Delta t_{GR}~~~~for~a~
graviton ,}
\end{equation}
then, the difference between the delay of neutrino and delay of
photon is given by
\begin{equation}
{(\Delta t_s)_\nu - (\Delta t_s)_\gamma = {\frac{1}{2}} \alpha \gamma \Delta t_{GR} .}
\end{equation}
Also the difference the delay of neutrino and delay of graviton is
given by
\begin{equation}
{(\Delta t_s)_\nu - (\Delta t_s)_g = {\frac{3}{2}} \alpha \gamma \Delta t_{GR} .}
\end{equation}

Now, taking $ (\alpha = {\frac{1}{137}}) $, $ \Delta t_{GR} = 12
\times 10^6  sec.$ as calculated before [8], for SN1987A, then
equation (9) gives $(t_1)_\nu$ then we get:
\begin{equation}
(t_1)_{\gamma} = (\Delta t_s)_\nu - (\Delta t_s)_\gamma = 4.4
\times 10^4 ~~sec.~~~~(for~~ \gamma = 1),
\end{equation}
 and since neutrinos from SN1987A arrived about $1.1 \times 10^4
sec. (= t_2)$ earlier than photons, thus the total time interval
between the emission of neutrinos and of photons, at the source,
is given by:
\begin{equation}
t_{\nu - \gamma} = t_1 + t_2 = 5.5 \times 10^4 sec. = 15^h.3~~~~
(for~ \gamma = 1)
\end{equation}
  So, neutrinos were emitted $15^h.3$ earlier than photons $(for~\gamma=1)$. Similarly, from
(10) we get, for gravitons,
\begin{equation}  (t_1)_g  = (\Delta
t_s)_{\nu} - (\Delta t_s)_{g} = 36^h.469~~~~~~(\gamma=1)
\end{equation}

and since neutrinos from SN1987A arrived $1.2~sec.$ later than
gravitons (about the same time) then the gravitons were emitted
$36^h.496 $ later than neutrinos $if \gamma=1$.

 The spin dependent
temporal model for SN1987A is now clear. Neutrinos were emitted
first, and after about $15^h.3$ photons were emitted signaling
maximum brightness of the envelope. After $36^h.5$, from the
emission of neutrinos (collapse of the core), gravitons were
emitted. This model is for $\gamma=1$.

\section{Discussion and Conclusion}

 SN observations provide us with the exact arrival times of different
types of carriers of information. Several authors have
investigated the arrival times of massless spinning particles
propagating through a background gravitational field (cf. [9]).
However, theoretical models are constructed to know the emission
times of such particles, knowing their observed arrival times. In
the present work, two schemes have been used to construct
chronological models for SN1987A. In the first we assume that
different carriers of information ( neutrinos, photons, gravitons)
follow the same spin independent trajectory (nul-geodesic) from
SN1987A to the Earth's detectors. In the second scheme we impose a
different assumption, that is the trajectories of different
carriers depend on the spin of the carrier (the new path equation
(2)). The resulting emission times of the three different carriers
are given in Table 1. 
\newpage
\begin{center}
 Table 1: Emission Times Given By The Models       \\
\vspace{0.5cm}
\begin{tabular}{|c|c|c|} \hline
& & \\ Particles Emitted  & Spin Independent Model  & Spin
Dependent Model
\\ & &
\\ \hline & & \\ Neutrino (core collapse)     & 0.0 & 0.0
\\ & & \\ \hline & & \\ Photons (maximum brightness) &  $ +3^h 5^m$   &
$+15^h 18^m$ \\ & & \\ \hline & & \\ Gravitons (?) &$-1^s.2$
&$+36^h 28^m$
\\ & & \\ \hline

\end{tabular}
\end{center}
\section*{}
As stated in section 2, the parameter $\gamma$ is a numerical
parameter of order unity. An upper limit for $\gamma(= 2)$ is
almost determined by comparing the results of applying (2) with
the experimental results of quantum interference of thermal
neutrons [10], known in the literature as the COW-experiment (cf.
[11], [12] and [13]). However, the spin dependent model of Table 1
is constructed using $\gamma = 1$. The time periods given by (11)
and (13), are doubled if $\gamma = 2$  and consequently the values
given in the last column of Table 1 will increase. The zero of the
time scale is adjusted at the time of emission of neutrinos (core
collapse). We prefer to leave the final value of the parameter
$(\gamma)$ to be fixed by future COW-experiment whose third
generation is in progress [12]. Depending on the quantities given
in the spin independent model of Table 1, which implies the
null-geodesic motion of different carriers, we can see that
neutrinos and gravitons are emitted at about the same time, while
photons are emitted $3^h 5^m$ later. If we assume that emission of
neutrinos is signaling  the collapse of the core, the emission of
gravitons is indicating the sudden change of space-time symmetry,
and the emission of photons is associated with the maximum
brightness of the envelope; then the collapse of the core of the
progenitor deviates from spherical symmetry and may give rise to a
kicked born neutron star [14] which changes the space-time
symmetry producing gravitational waves. While the maximum
brightness of the envelope is achieved after about three hours
from the collapse of the core. This is the model obtained from
using the spin independent assumption.

Using the results given in the spin dependent model of Table 1,
one can see that photons are emitted after about $15^h$ from the
collapse of the core. While gravitons are emitted after about
$36^h$ from the collapse of the core, or $21^h$ after the phase of
the maximum brightness. This means that space-time symmetry is
changed after $36^h$ from the collapse of the core. Then we can
conclude that the collapse was spherically symmetric, and also,
the expansion till the maximum brightness. In this case, we can
relate the emission of gravitational waves to an asymmetric
explosion happened after $21^h$ from the maximum brightness phase.

To distinguish between the two models we consider the following
points, arising from the accepted mechanism of SN: \\ 1. Mechanism
for type II supernovae suggest that neutrinos are emitted tens of
hours earlier than photons [15]. \\ 2. There is a general
agreement that the collapse of the core is symmetric. This is
because neutrino radiation damps out the non-sphericity of the
collapse [16].

Using these two  criteria we may conclude that the spin dependent
model is more accepted than the spin independent one.

\section*{References}
 {1.} D. N. Schramm and J. W. Truran,   {\it Physics
Reports}, {\bf 189}, 89(1990). \\ {2.} J. Weber, Proc. First
Edoardo Amaldi Conf.

  Ed.
E.Coccia et al. World Scientific, 416(1994).
\\
{3.} A. De Rujula, {\it Phys. Lett.}, {\bf 60}, 176(1987).\\
 {4.} M. I. Wanas, M. Melek  and M. E.
Kahil,   {\it Astrophys. Space Sci.} {\bf 228}, 273(1995).
\\{5.} M. I. Wanas,  {\it Astrophys. Space Sci.}, {\bf
258}, 237(1998). \\{6.} L. M. Krauss,  and S. Tremaine,   {\it
Phys. Rev. Lett.}, {\bf 60}, 176(1988). \\ {7.} M. J. Longo, {\it
Phys. Rev. D}, {\bf 36}, 3276(1987). \\{8.} M. J. Longo, {\it
Phys. Rev. Lett.}, {\bf 60}, 173(1988).\\{9.} L. Halpern, Proc.
Fifth Marcel Grossmann Meeting  , Ed. D.G.Blair,

M.J. Buckingham and R. Ruffini, 1143(1988).\\ {10.} M. I. Wanas,
M. Melek and M. E. Kahil,  gr-qc/9812085(1998) . \\ {11.}
 S. A. Werner, H. Kaiser, M. Arif and R. Clothier,  {\it
Physica B},{\bf 151}, 22(1988). {12.} M. Arif, M. S. Dewey, G. L.
Greene, D.  Jacobson and S. Werner, {\it Phys.}

 {\it
Lett.A}, {\bf 184} ,154(1994). \\ {13.} K. C. Littrell, B.C.
Allman and S. A. Werner,   {\it Phys. Rev.A }, {\bf 56},
1767(1997).\\ {14.} K. Kohri and S.Nagantaki,
astro-ph/9911077(1999).\\
 {15.} A. De Rujula,  Preprint
CERN-TH. 5012/88(1988). \\{16.} D. Kazanas and D. Schramm, {\it
Astrophys. J.}, {\bf 214}, 819(1976).

\end{document}